\pdfoutput=1
\documentclass[conference,letterpaper,10pt]{IEEEtran}

\usepackage[latin1]{inputenc}
\usepackage{times,amsmath}
\usepackage{amsfonts}
\usepackage{pstool}
\usepackage{subfigure}
\usepackage{multirow}
\usepackage{enumerate}
\usepackage{graphicx}
\usepackage{MnSymbol}
\usepackage{stfloats} 
\usepackage[table]{xcolor}
\usepackage[square, comma, sort&compress, numbers]{natbib}
\usepackage{nohyperref}
\usepackage{algorithm,algorithmic}
\usepackage{bigints}

\newcounter{tempEquationCounter} 
\newcounter{thisEquationNumber}

\makeatletter
\newcommand{\vast}{\bBigg@{4}}
\newcommand{\Vast}{\bBigg@{5}}
\makeatother

\graphicspath{ {Figures/} }

\begin{document}

\title{ Power Imbalance Detection in Smart Grid via Grid Frequency Deviations: A Hidden Markov Model based Approach  }

\author{
\IEEEauthorblockN{ Shah Hassan\IEEEauthorrefmark{1}, Hadia Sajjad\IEEEauthorrefmark{1}, Muhammad Mahboob Ur Rahman\IEEEauthorrefmark{1} }
\IEEEauthorblockA{\IEEEauthorrefmark{1} Electrical engineering department, Information Technology University, Lahore, Pakistan \\\{hadia.sajjad,mahboob.rahman\}@itu.edu.pk }
}

\maketitle

\begin{abstract} 

We detect the deviation of the grid frequency from the nominal value (i.e., 50 Hz), which itself is an indicator of the power imbalance (i.e., mismatch between power generation and load demand). We first pass the noisy estimates of grid frequency through a hypothesis test which decides whether there is no deviation, positive deviation, or negative deviation from the nominal value. The hypothesis testing incurs miss-classification errors---false alarms (i.e., there is no deviation but we declare a positive/negative deviation), and missed detections (i.e., there is a positive/negative deviation but we declare no deviation). Therefore, to improve further upon the performance of the hypothesis test, we represent the grid frequency's fluctuations over time as a discrete-time hidden Markov model (HMM). We note that the outcomes of the hypothesis test are actually the emitted symbols, which are related to the true states via emission probability matrix. We then estimate the hidden Markov sequence (the true values of the grid frequency) via maximum likelihood method by passing the observed/emitted symbols through the Viterbi decoder. Simulations results show that the mean accuracy of Viterbi algorithm is at least $5$\% greater than that of hypothesis test.

\end{abstract}


\section{Introduction}
\label{sec:intro}

Fluctuations of the instantaneous grid frequency are considered to be a viable indicator of the power imbalance (disparity between the generation and load demand) in a grid. For example, negative frequency deviation hints at sudden shortfall of generation which could lead to a potential outage/blackout. Traditionally, it is the generation side which is responsible to restore the grid frequency to its nominal value in a short time. To this end, the generators measure and keep track of the instant grid frequency, and employ the so-called {\it frequency control} to bring the grid frequency back to nominal value when a (positive or negative) frequency deviation greater than a threshold is detected. 

Specifically, the traditional (generation-side) approach to ensure grid frequency stability consists of three levels/tiers of frequency control which are triggered on different time-scales \cite{Yu:MIT:1996}. The primary control, commonly known as frequency response (FR), is activated within few seconds after the disturbance. FR instantly adjusts the governors (i.e., the speed of the motors) on the generation side to increase/decrease the generation power to stabilize the instant grid frequency. The secondary control, commonly known as load frequency control (LFC) or automatic generation control (AGC), consists of both spinning and non-spinning reserves which are utilized to adjust the generated power on need basis \cite{Jaleeli:TPS:1992}. LFC stablizes grid frequency on a time-scale of minutes. The tertiary control, known as economic dispatch (ED), ensures the stability of the grid frequency by changing the set-points of each of the generators to meet the current load demand at minimum operating cost \cite{Raghu:TPS:2012}, \cite{Yasmeen:ACC:2012}.

The extravagant cost of generation-side frequency control (due to spinning reserves) has prompted interest in load-side (demand response based) frequency control whereby the consumer load switches on and off to adjust its instant load demand after observing a frequency deviation (see \cite{Moghadam:TSG:2014},\cite{Short:TPS:2007},\cite{Molina:TPS:2011} and the reference therein). Load-side frequency control, previously considered to be infeasible, is now considered to be a viable solution to grid frequency stability, thanks to the bi-directional signalling (between generation side and the load side) enabled by the smart grid.

The performance of the generation-side/load-side frequency control (especially, the economic dispatch problem) relies critically upon the grid frequency measurements\footnote{The grid frequency measurements have traditionally been collected by a (synchrophasor based) phasor measurement unit (PMU), or, more recently, via a frequency disturbance recorder (FDR).}, which are noisy, and thus, not reliable on their own. The crux of this work, therefore, is to represent the true grid frequency time-series as a hidden Markov model (HMM). This enables us to extract via Viterbi algorithm the true states of HMM, given a sequence of noisy measurements. The output of the Viterbi algorithm could then be utilized by the frequency control mechanism to ensure grid frequency stability. The main contributions of this work are formally summarized below:
\begin{itemize} 
\item
We represent the true grid frequency time-series as a hidden Markov model. This enables us to extract via Viterbi algorithm the true states of HMM, given a sequence of noisy measurements of instant grid frequency.

\item
We carry out hypothesis testing on noisy measurements of grid frequency to generate the emitted symbols (i.e., the entries of the emission probability matrix) for the HMM.
\end{itemize}

{\bf Outline.} The rest of this paper is organized as follows. Section-II introduces the system model. The hypothesis testing framework to generate the emitted symbols is described in section-III. In section-IV, we model the time-evolution of true grid frequency via a hidden Markov chain, and discover the hidden states via Viterbi algorithm. Section-V provides numerical results followed by discussions. Finally, Section-VI concludes the paper.

\section{System Model \& Background}
\label{sec:sys-model}

\subsection{System Model}
\begin{figure}[ht]
\begin{center}
	\includegraphics[width=2.8in]{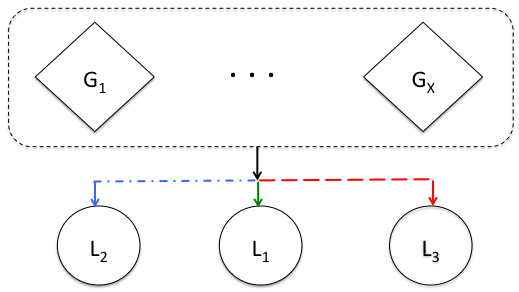} 
\caption{The system model: $X$ generators generate a cumulative power $P_G[k]$ to potentially serve a load consisting of three discrete load elements ($L_1,L_2,L_3$) with cumulative power $P_L[k]$. By default, the generation caters for the medium load (say, when $L_1$ and $L_2$ are active), leading to zero power imbalance, i.e., $P_G[k]=P_L[k]$. In this case, no frequency deviation is detected on the grid. On the other hand, the scenario of light (heavy) load, say, when $L_2$ is switched off ($L_3$ is switched on) leads to positive (negative) power imbalance, i.e., $P_G>P_L$ ($P_G<P_L$). In such situation, positive (negative) frequency deviation is detected on the grid. }
\label{fig:sys-model}
\end{center}
\end{figure}

In this work, we consider a simplified model for smart grid. Specifically, there are $X$ number of power generators which together generate a power $P_G[k]$ at time $k$; the load demand at time $k$ is represented by $P_L[k]$. We further assume that the system's load consists of three discrete elements; therefore, the system is in one of the three states at time $k$: low load, medium load, heavy load (see Fig. \ref{fig:sys-model}).  

\subsection{Background: Frequency Deviation for Power Imbalance Detection}
Let $f[k]$ represent the instantaneous grid frequency, while $f_0=50$ Hz is the nominal grid frequency. Let $\Delta P[k]=P_G[k]-P_L[k]$ denote the instant power imbalance, and $\Delta f[k]=f[k]-f_0$ denote the instant frequency deviation. Then, it is well-known in the literature that frequency deviation is a monotonic function of power imbalance (i.e., it is difficult to characterize the exact analytical relationship between the two quantities) \cite{Raghu:TPS:2012},\cite{Basu:ISTA:2016}. In other words, we have: $\Delta f[k] \propto \Delta P[k]$. This relationship is known as power-frequency characteristic in the literature. 

\begin{figure}[ht]
\begin{center}
	\includegraphics[width=3.5in]{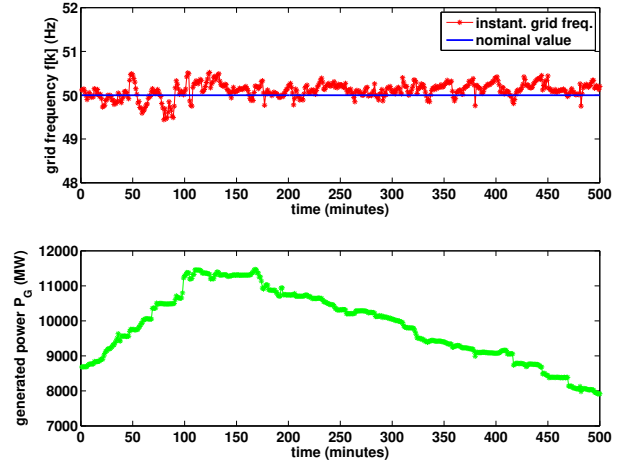} 
\caption{Real-time, time-series of grid frequency and generated power: The data corresponds to Pakistan national grid [Courtesy: NPCC, Pakistan]. }
\label{fig:fts}
\end{center}
\end{figure}

Fig. \ref{fig:fts} plots ($\sim 8$ hours) time-series of measurements of grid frequency and generated power respectively. The data for Fig. \ref{fig:fts} corresponds to the national grid of Pakistan, and was provided by National Power Control Center (NPCC), Pakistan. Fig. \ref{fig:fts} shows that the frequency deviation stays mostly negative (positive) for first 100 (last 400 minutes) minutes; therefore, the frequency control mechanism of the grid keeps increasing (decreasing) the generated power during the same interval to ensure frequency stability. In short, Fig. \ref{fig:fts} attests to the fact that the instant frequency deviation is indeed a viable indicator of the current power imbalance in the system.

\section{Power Imbalance Detection via Hypothesis Testing}

Let $\Delta f_{max}$, $\Delta f_{min}$ represent the maximum and minimum frequency deviation from the nominal value. $\Delta f_{max}$, $\Delta f_{min}$ are typically estimated from historical data. Let $z[k]$ denote the noisy measurement of grid frequency at time $k$.

\begin{figure}[ht]
\begin{center}
	\includegraphics[width=3.5in]{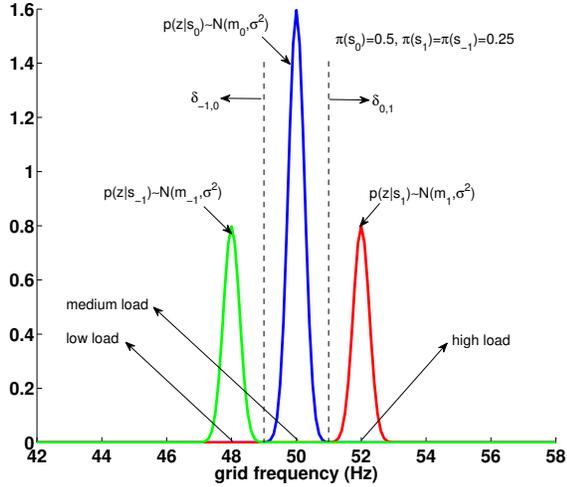} 
\caption{The inefficient/slow frequency control could lead to a situation where the distribution/histogram of the (measured) grid frequency time-series $\{z[k]\}$ becomes multi-modal with small side-lobes. This picture illustrates by example that for the system model considered in Fig. 1, $\{z[k]\}$ could be multi-modal with three lobes.}
\label{fig:HT}
\end{center}
\end{figure}

At a given time instant $k$, w.r.t. the grid frequency the smart grid is in one of the three states with the state-space: $\mathcal{S}=\{{-1},0,1\}$ (see Fig. \ref{fig:HT}). The states $s{[k]}={-1},s{[k]}=0,s{[k]}=1$ imply that the frequency deviation is negative, zero, positive respectively at time $k$\footnote{This could also be represented by a Gaussian mixture model (to model the multi-modal nature of the distribution of the measurement $z$).}. The no frequency deviation implies no power imbalance, positive frequency deviation implies excessive generation, and negative frequency deviation implies shortfall of generation. Therefore, the same state-space model holds for both frequency deviation and power imbalance.

Assuming Gaussian measurement error, the maximum-likelihood (ML) test boils down to the following distance-based hypothesis test:
\begin{equation}
x[k]=\delta_{ML}(z[k])= \arg \min_{i \in \{ -1,0,1 \}} | z[k] - m_i | 
\end{equation} 
where $m_{-1}=f_0-\Delta f_{min}$ Hz, $m_0=f_0$ Hz, $m_1=f_0+\Delta f_{max}$ Hz. With Gaussian measurement error, $z[k]|(s{[k]}=i)\sim N(m_i,\sigma^2)$, where $\sigma^2$ is the variance of the estimation error. This gives the following ML decision rule: 
\begin{equation}
x[k] = \left\{\begin{array}{lr}
        -1, & \text{for } z \leq \delta_{-1,0} \\
        0, & \text{for } \delta_{-1,0} \leq z \leq \delta_{0,1} \\
        1, & \text{for } \delta_{0,1} \leq z
        \end{array}\right\}
\end{equation}
Let $P(s[0]=-1)=\pi(-1); P(s[0]=0)=\pi(0);P(s[0]=1)=\pi(1)$ be the prior probabilities for the three states. Then, $\delta_{-1,0}=\frac{m_{-1}+m_0}{2}+\frac{\eta_{-1,0} \sigma^2}{m_{0}-m_{-1}}$, and $\delta_{0,1}=\frac{m_0+m_1}{2}+\frac{\eta_{0,1} \sigma^2}{m_{1}-m_0}$, where $\eta_{-1,0}=\log_e (\pi(-1)/\pi(0))$ and $\eta_{0,1}=\log_e (\pi(0)/\pi(1))$. For the special case of equal priors, $\delta_{-1,0}=\frac{m_{-1}+m_0}{2}$, and $\delta_{0,1}=\frac{m_0+m_1}{2}$. Then, we have the following expressions for the error probabilities, i.e., deciding state $j$ though the true state was $i$:
\begin{equation}
\begin{split}
P_{e,-1} = P_e |(s[k]=-1) &=\sum_{j \in \{0,1\}} P(x[k]=j|s[k]=-1) \\
& = Q(\frac{\delta_{-1,0}-m_{-1}}{\sigma})
\end{split}
\end{equation}
where $Q(x)=\frac{1}{\sqrt{2\pi}} \int_x^\infty  e^{-\frac{t^2}{2}} dt$ is the standard $Q$-function. Similarly, 
\begin{equation}
\begin{split}
P_{e,0} = P_e |(s[k]=0) &=\sum_{j \in \{-1,1\}} P(x[k]=j|s[k]=0) \\
& = 1 - \bigg( Q(\frac{\delta_{-1,0}-m_{0}}{\sigma})-Q(\frac{\delta_{0,1}-m_{0}}{\sigma}) \bigg)
\end{split}
\end{equation}
Finally, 
\begin{equation}
\begin{split}
P_{e,1} = P_e |(s[k]=1) &=\sum_{j \in \{-1,0\}} P(x[k]=j|s[k]=1) \\
& = 1 - Q(\frac{\delta_{0,1}-m_{1}}{\sigma})
\end{split}
\end{equation}

Then, one can define $P_{d,-1}=1-P_{e,-1}$, $P_{d,0}=1-P_{e,0}$, $P_{d,1}=1-P_{e,1}$ as the probability of correctly detecting that the system is in state $-1$, $0$, $1$, respectively.

\section{Hidden Markov Model for Grid Frequency Evolution over Time}
\label{sec:EH}

\subsection{The Hidden Markov Model}
As briefly mentioned in Section-III, the true grid frequency remains always in one of the three states with the state-space: $\mathcal{S}=\{{-1},0,1\}$. The states $s{[k]}={-1},s{[k]}=0,s{[k]}=1$ imply that the frequency deviation is negative, zero, positive respectively at time $k$. But $\{s[k]\}$ constitutes a Markov chain which is hidden; therefore, $\{x[k]\}$ that we observe through the hypothesis test are the so-called emitted symbols. The connection between the true/hidden states and the emitted symbols is given by the  emission probability matrix: 
\begin{equation}
\mathbf{R} = 
\begin{bmatrix}
  r_{-1,-1} & r_{-1,0} & r_{-1,1} \\
  r_{0,-1} & r_{0,0} & r_{0,1} \\
  r_{1,-1} & r_{1,0} & r_{1,1}
 \end{bmatrix}
\end{equation}
where $r_{i,j}=P(x[k]=i|s[k]=j)$, $i,j \in \{ -1,0,1 \}$. One can verify that the sum of the elements in each column of $\mathbf{R}$ is 1. The off-diagonal elements in the $i$-th row of $\mathbf{R}$ represent the errors made by the ML test, i.e., deciding the state as $x[k]=i$, $i \in \{-1,0,1\}\setminus{j}$ while the system was actually in state $s[k]=j$. The elements $r_{i,j}$ of the matrix $\mathbf{R}$ are listed in Table \ref{tb:R}.

\begin{table}[h]
\begin{center}
    \begin{tabular}{ | l | p{5cm} |}
    \hline
    probability & expression \\ \hline
    $r_{-1,-1}$ & $1-Q(\frac{\delta_{-1,0}-m_{-1}}{\sigma})$ \\ \hline
    $r_{-1,0}$ & $1-Q(\frac{\delta_{-1,0}-m_{0}}{\sigma})$ \\ \hline
    $r_{-1,1}$ & $1-Q(\frac{\delta_{-1,0}-m_{1}}{\sigma})$ \\ \hline
    $r_{0,-1}$ & $Q(\frac{\delta_{-1,0}-m_{-1}}{\sigma}) - Q(\frac{\delta_{0,1}-m_{-1}}{\sigma})$ \\ \hline
    $r_{0,0}$ & $Q(\frac{\delta_{-1,0}-m_{0}}{\sigma})-Q(\frac{\delta_{0,1}-m_{0}}{\sigma})$ \\ \hline
    $r_{0,1}$ & $Q(\frac{\delta_{-1,0}-m_{1}}{\sigma})-Q(\frac{\delta_{0,1}-m_{1}}{\sigma})$ \\ \hline
    $r_{1,-1}$ & $Q(\frac{\delta_{0,1}-m_{-1}}{\sigma})$ \\ \hline
    $r_{1,0}$ & $Q(\frac{\delta_{0,1}-m_{0}}{\sigma})$ \\ \hline
    $r_{1,1}$ & $Q(\frac{\delta_{0,1}-m_{1}}{\sigma})$ \\ \hline
    \end{tabular}
    \caption{The entries of the emission probability matrix $\mathbf{R}$.}
    \label{tb:R}
\end{center}
\end{table}

\begin{figure}[ht]
\begin{center}
	\includegraphics[width=3.5in]{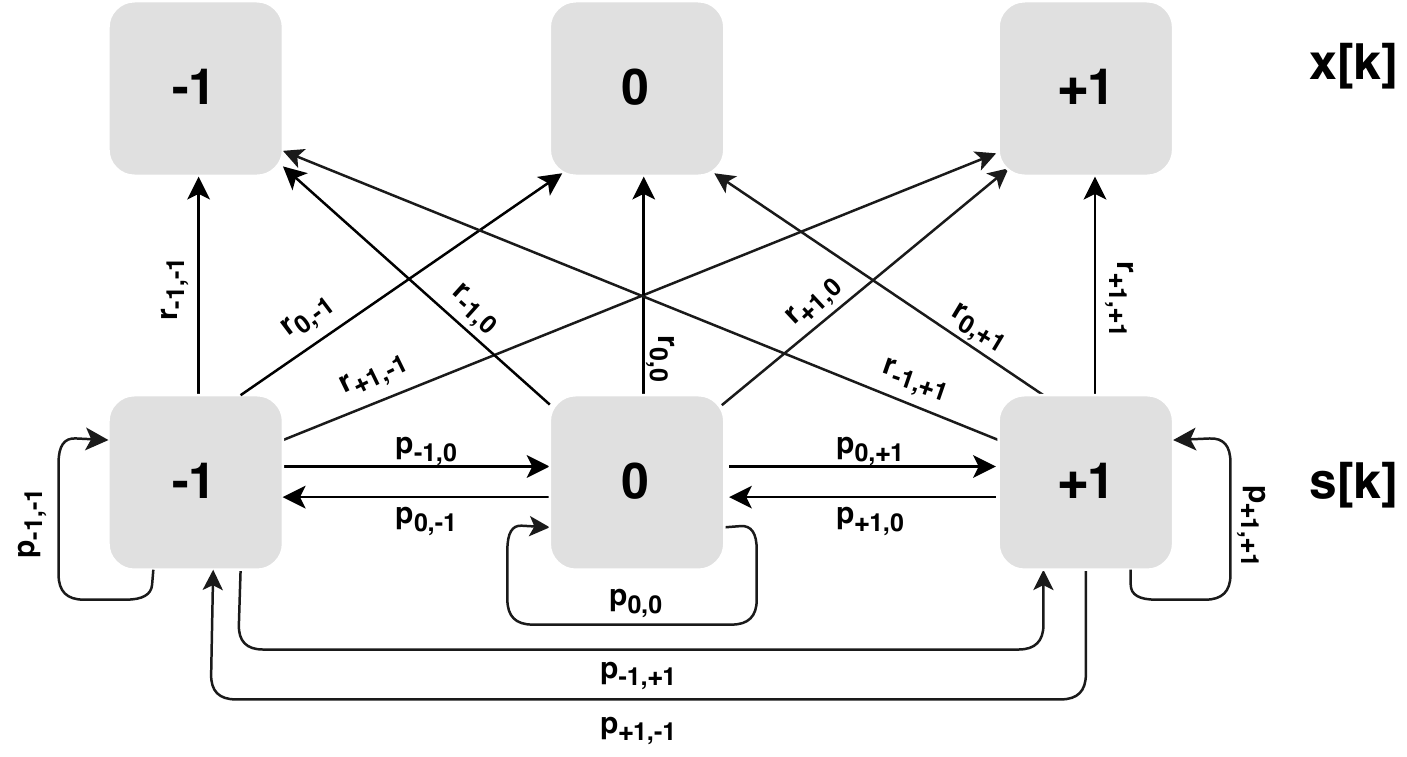} 
\caption{The hidden Markov model for the grid frequency time-evolution: $p_{i,j}$($r_{i,j}$) represent the entries of transition (emission) probability matrix.}
\label{fig:MC}
\end{center}
\end{figure}

The transition from state $s[k-1]=i$ to state $s[k]=j$ occurs after a fixed interval of $T=t_k-t_{k-1}$ seconds where $1/T$ is the measurement rate (of grid frequency). To this end, we have the following transition probability matrix for the hidden Markov chain:
\begin{equation}
\mathbf{P} = 
\begin{bmatrix}
  p_{-1,-1} & p_{-1,0} & p_{-1,1} \\
  p_{0,-1} & p_{0,0} & p_{0,1} \\
  p_{1,-1} & p_{1,0} & p_{1,1}
 \end{bmatrix}
\end{equation}
where $p_{i,j}=P(s[k]=j|s[k-1]=i)$, $i,j \in \{ -1,0,1 \}$. One can verify that the sum of the elements in each row of $\mathbf{P}$ is 1. Fig. \ref{fig:MC} provides a graphical summary of the essentials of HMM considered, i.e., the hidden markov chain $\{s[k]\}$, time evolution $\mathbf{P}$ of $\{s[k]\}$, the emitted symbols $\{x[k]\}$, and the connection $\mathbf{R}$ between $\{s[k]\}$ and $\{x[k]\}$. 

\subsection{Maximum likelihood Estimation of Hidden Markov sequence via Viterbi Algorithm}
We utilize Viterbi algorithm to obtain maximum likelihood sequence estimate (MLSE) $\{s^*[k]\}_{k=1}^K = \mathbf{s^*}$ of $\{s[k]\}_{k=1}^K = \mathbf{s}$, given $\{x[k]\}_{k=1}^K = \mathbf{x}$ as follows:
\begin{equation}
\mathbf{s^*} = \arg \max_{\mathbf{s^{'}}} P(\mathbf{x},\mathbf{s^{'}}) = \arg \max_{\mathbf{s^{'}}} P(\mathbf{x}|\mathbf{s^{'}}) P(\mathbf{s^{'}})
\end{equation}
where
\begin{equation}
P(\mathbf{x}|\mathbf{s}) = \prod_{k=1}^K P(x[k]=x_k|s[k]=s_k) = \prod_{k=1}^K r_{k,k} 
\end{equation}
and
\begin{equation}
\begin{split}
P(\mathbf{s}) &= P(s[1]=s_1)\prod_{k=1}^K P(s[k+1]=s_{k+1}|s[k]=s_k) \\ 
&= P(s[1]=s_1)\prod_{k=1}^K p_{k,k+1}
\end{split}
\end{equation}

Therefore, we obtain the following expression for the joint probability $P(\mathbf{x},\mathbf{s})$:
\begin{equation}
\label{eq:JP2}
\begin{split}
P(\mathbf{x},\mathbf{s}) &= P(s[1]=s_1) \prod_{k=1}^K r_{k,k} p_{k,k+1}
\end{split}
\end{equation}
The joint probability of Eq. (\ref{eq:JP2}) is still hard to compute. To this end, Viterbi algorithm utilizes dynamic programming approach to break this problem into smaller sub-problems via a recursive (Trellis-based) approach.

\section{Numerical Results and Discussions}
\label{sec:results}

\subsection{Simulation Setup}
Let $\Pi_0=[\pi(-1), \pi(0), \pi(1)]'$. Then, one can see that the computation of MLSE depends upon knowledge of $\mathbf{P}$, $\mathbf{R}$ and $\Pi_0$. We consider the following transition probability matrix:
\begin{equation}
\mathbf{P} = 
\begin{bmatrix}
  0.2 & 0.7 & 0.1 \\
  0.1 & 0.8 & 0.1 \\
  0.1 & 0.7 & 0.2
 \end{bmatrix}
\end{equation}
where the entries are designed such that all the transitions to (away from) state 0 are very likely (unlikely). In other words, whenever there is frequency deviation (and thus, power imbalance), it is quickly eradicated to zero, thanks to the tertiary frequency control mechanism employed by the smart grid.  

The emission probability matrix depend upon a number of parameters (see Table-I). For example, with $\sigma=0.2$ Hz, $\pi (-1) = 0.1$, $\pi (0) = 0.8$, $\pi (1) = 0.1$, $m_{-1} = 49$ Hz, $m_0 = 50$ Hz, $m_1 = 51$ Hz, one gets the following $\mathbf{R}$:  
\begin{equation}
\label{eq:Rnum}
\mathbf{R} = 
\begin{bmatrix}
  	0.9814  &  0.0018  &  0.0000 \\
    0.0186  &  0.9965  &  0.0186 \\
    0.0000  &  0.0018  &  0.9814
 \end{bmatrix}
\end{equation}

\subsection{Simulation Results}

\begin{figure}[ht]
\begin{center}
	\includegraphics[width=3.5in]{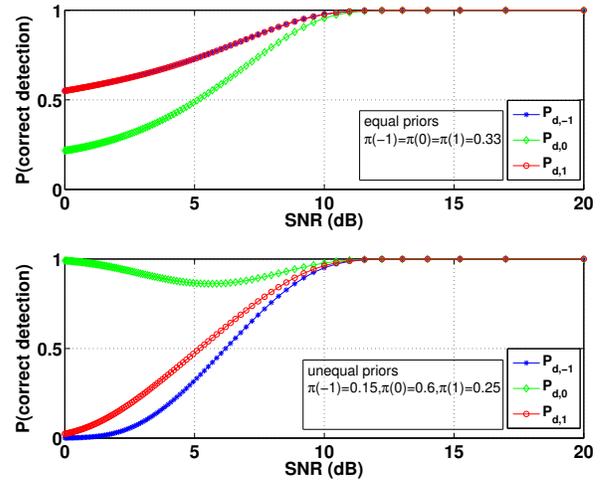} 
\caption{probabilities of correct detection vs. SNR}
\label{fig:Pd}
\end{center}
\end{figure}

Fig. \ref{fig:Pd} plots the probabilities of correct detection against signal-to-noise ratio (SNR) where the SNR (quality of measurements) is defined as $1/\sigma$. Specifically, the top figure considers the case of equal priors, i.e., $\pi(-1)=0.33, \pi(0)=0.34, \pi(1)=0.33$, while the bottom figure considers the case of unequal priors i.e., $\pi(-1)=0.15,\pi(0)=0.6,\pi(1)=0.25$. Also, to obtain Fig. \ref{fig:Pd}, we have set $m_{-1} = 49.6$ Hz; $m_0 = 50$ Hz; $m_1 = 50.4$ Hz. Fig. \ref{fig:Pd} reveals that there is a threshold SNR ($\sim 12$ dB in the case) exceeding which implies that $P(x[k]=s[k]) \to 1$. In other words, as the measurements become more and more reliable, the hidden markov chain starts to become more and more visible, and vice versa.

\begin{figure}[ht]
\begin{center}
	\includegraphics[width=3.5in]{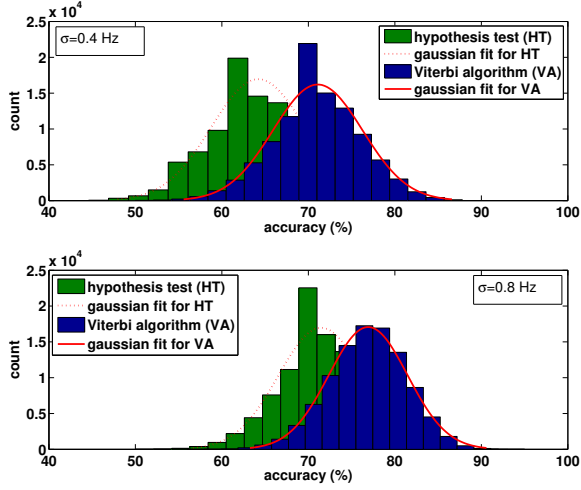} 
\caption{Accuracy of hypothesis test and Viterbi algorithm, for $\sigma=0.4$ Hz, $\sigma=0.8$ Hz respectively. }
\label{fig:VAA}
\end{center}
\end{figure}

Fig. \ref{fig:VAA} is the histogram plot of the accuracy of hypothesis test (HT) and Viterbi algorithm (VA) (i.e., the distance of outcomes of the HT and VA from the sequence of the hidden states $\{s[k]\}$. For this plot, we first generated a length-$K=100$ sequence $\{s[k]\}_{k=1}^{K=100}$ modelling the true/hidden states. We then generated a length-100 sequence $\{x[k]\}_{k=1}^{K=100}$ of the observed/emitted symbols (using matrix $\mathbf{R}$), which represents the outcome of the hypothesis test as well. We then applied Viterbi algorithm to do MLSE, i.e., to compute $\{s^*[k]\}_{k=1}^{K=100}$. We computed the accuracy of hypothesis test as: $\#(x[k]\neq s[k])/K$, and of Viterbi algorithm as: $\#(s^*[k]\neq s[k])/K$. We repeated the same procedure 100,000 times (as per Monte-Carlo simulations methodology) to get meaningful and reliable (average) results. Furthermore, we set $\pi(-1)=0.25,\pi(0)=0.6,\pi(1)=0.15$; $m_{-1} = 49.4$ Hz; $m_0 = 50$ Hz; $m_1 = 50.7$ Hz.

The top (bottom) figure of Fig. \ref{fig:VAA} plots the histograms representing the accuracy of the hypothesis test and Viterbi algorithm for $\sigma=0.4$ Hz ($\sigma=0.8$ Hz). For each of the two figures, each red curve represents the best Gaussian fit to the respective histogram. From Fig. \ref{fig:VAA}, one easily infer that Viterbi algorithm indeed performs better than the hypothesis test. Specifically, with the Gaussian approximation, we have: $\mu_{HT} = 64.1998$, $\sigma_{HT} = 5.41401$, and $\mu_{VA} = 71.0862$, $\sigma_{VA} =  5.1705$ for the top figure. While we have: $\mu_{HT} = 71.4686$, $\sigma_{HT} = 4.93606$, and $\mu_{VA} = 76.9156$, $\sigma_{VA} = 4.55722$ for the bottom figure. Thus, the mean accuracy of Viterbi algorithm is at least $5$\% greater than that of hypothesis test. Therefore, the Viterbi algorithm could be thought of as a filter which takes at its input a noisy sequence $\{x[k]\}$, and returns at its output a cleaner sequence $\{s^*[k]\}$.  

Last but not the least, to our surprise, mean accuracy of both the hypothesis test and the Viterbi algorithm increases with the increase in standard deviation of the measurement error.

\subsection{Discussions}
\begin{itemize}
\item
The proposed HMM framework could be used for ($m>0$-step ahead) prediction of frequency deviation/power imbalance as follows. Let $\vec{s}[k]=[P(s[k]=-1), P(s[k]=0), P(s[k]=1)]^T$. Assuming that the system was in state $0$ at time $k=0$, i.e. $\vec{s}[0]=[0, 1, 0]^T$ and we are in time $k-1$, we want to predict the probability vector $\vec{s}[k]$ at time $k$ that the system is in state $i$, $i \in \{-1,0,1\}$. Then, we have the following recursive relation: $\vec{s}[k]=\mathbf{P}\vec{s}[k-1]$. Alternatively, we can write: $\vec{s}[k]=\mathbf{P}^k\vec{s}[0]$.

\item
Baum-Welch/forward-backward algorithm could be used to systematically learn the HMM paramters ($\mathbf{P}$,$\mathbf{R}$,$\Pi_0$) from the measured data.
\end{itemize}

\section{Conclusion}
\label{sec:conclusion}

In this preliminary work, we represented the grid frequency's fluctuations over time as a discrete-time hidden Markov model. The emitted symbols for the considered HMM were obtained by carrying out hypothesis testing on the noisy measurements of grid frequency. We then recovered the hidden markov sequence (of true grid frequency values) by maximum likelihood method by passing the emitted symbols through the Viterbi decoder. Simulations results showed that the mean accuracy of Viterbi algorithm is at least $5$\% greater than that of hypothesis test.

Immediate future work will look into the following: i) HMM-enabled prediction of power imbalance (and potential outage), ii) implementation of the Baum-Welch/forward-backward algorithm to estimate/learn the parameters of HMM from the measured (but unlabelled) data, iii) implementation of the proposed HMM based framework for power imbalance detection in a more sophisticated and realistic system, e.g., IEEE 13-bus system, and iv) quantitative evaluation of the (positive) impact of the proposed HMM based framework on the performance of the tertiary control, i.e., economic dispatch.

\appendices


\footnotesize{
\bibliographystyle{IEEEtran}
\bibliography{references}

\begin{thebibliography}{1}
\providecommand{\url}[1]{#1}
\csname url@rmstyle\endcsname
\providecommand{\newblock}{\relax}
\providecommand{\bibinfo}[2]{#2}
\providecommand\BIBentrySTDinterwordspacing{\spaceskip=0pt\relax}
\providecommand\BIBentryALTinterwordstretchfactor{4}
\providecommand\BIBentryALTinterwordspacing{\spaceskip=\fontdimen2\font plus
\BIBentryALTinterwordstretchfactor\fontdimen3\font minus
  \fontdimen4\font\relax}
\providecommand\BIBforeignlanguage[2]{{%
\expandafter\ifx\csname l@#1\endcsname\relax
\typeout{** WARNING: IEEEtran.bst: No hyphenation pattern has been}%
\typeout{** loaded for the language `#1'. Using the pattern for}%
\typeout{** the default language instead.}%
\else
\language=\csname l@#1\endcsname
\fi
#2}}

\bibitem{Yu:MIT:1996}
C.-N. Yu, ``Real power and frequency control of large electric power systems
  under open access,'' Ph.D. dissertation, Massachusetts Institute of
  Technology, 1996.

\bibitem{Jaleeli:TPS:1992}
N.~Jaleeli, L.~S. VanSlyck, D.~N. Ewart, L.~H. Fink, and A.~G. Hoffmann,
  ``Understanding automatic generation control,'' \emph{IEEE Transactions on
  Power Systems}, vol.~7, no.~3, pp. 1106--1122, Aug 1992.

\bibitem{Raghu:TPS:2012}
R.~Mudumbai, S.~Dasgupta, and B.~B. Cho, ``Distributed control for optimal
  economic dispatch of a network of heterogeneous power generators,''
  \emph{IEEE Transactions on Power Systems}, vol.~27, no.~4, pp. 1750--1760,
  Nov 2012.

\bibitem{Yasmeen:ACC:2012}
A.~Yasmeen, R.~Mudumbai, and S.~Dasgupta, ``A distributed algorithm for optimal
  dispatch in smart power grids with piecewise linear cost functions,'' in
  \emph{2012 2nd Australian Control Conference}, Nov 2012, pp. 36--40.

\bibitem{Moghadam:TSG:2014}
M.~R.~V. Moghadam, R.~T.~B. Ma, and R.~Zhang, ``Distributed frequency control
  in smart grids via randomized demand response,'' \emph{IEEE Transactions on
  Smart Grid}, vol.~5, no.~6, pp. 2798--2809, Nov 2014.

\bibitem{Short:TPS:2007}
J.~A. Short, D.~G. Infield, and L.~L. Freris, ``Stabilization of grid frequency
  through dynamic demand control,'' \emph{IEEE Transactions on Power Systems},
  vol.~22, no.~3, pp. 1284--1293, Aug 2007.

\bibitem{Molina:TPS:2011}
A.~Molina-Garcia, F.~Bouffard, and D.~S. Kirschen, ``Decentralized demand-side
  contribution to primary frequency control,'' \emph{IEEE Transactions on Power
  Systems}, vol.~26, no.~1, pp. 411--419, Feb 2011.

\bibitem{Basu:ISTA:2016}
M.~Basu, R.~Mudumbai, and S.~Dasgupta, ``Intelligent distributed economic
  dispatch in smart grids,'' in \emph{Intelligent Systems Technologies and
  Applications}.\hskip 1em plus 0.5em minus 0.4em\relax Springer, 2016, pp.
  285--295.

\end{thebibliography}
}

\vfill\break

\end{document}